\documentclass[aps,prl,twocolumn]{revtex4-1}

\usepackage{graphics}      
\usepackage{graphicx}      
\usepackage{longtable}     
\usepackage{url}           
\usepackage{bm,color}            
\usepackage{amssymb, amsmath, amsthm, enumerate, epsfig,subfigure,setspace}
\usepackage{multirow}

\theoremstyle{plain}
\newtheorem*{prop*}{Proposition}

\definecolor{red}{rgb}{0.8, 0.25, 0.33}
\definecolor{green}{rgb}{0.0, 0.5, 0.0}

\begin{document}

\title{Large-scale Sustainable Search on Unconventional Computing Hardware}

\author{Kirill P. Kalinin${}^{1\dagger}$ and Natalia G. Berloff${}^{2,1}$}

\email[correspondence address: ]{kpk26@cam.ac.uk, N.G.Berloff@damtp.cam.ac.uk}
\affiliation{${}^1$Department of Applied Mathematics and Theoretical Physics, University of Cambridge, Cambridge CB3 0WA, United Kingdom}
\affiliation{${}^2$Skolkovo Institute of Science and Technology,  Bolshoy Boulevard 30, build. 1 Moscow, 121205 Russian Federation}

\date{\today}

\begin{abstract}{
Since the advent of the Internet, quantifying the relative importance of web pages is at the core of search engine methods. According to one algorithm, PageRank, the worldwide web structure is represented by the Google matrix, whose principal eigenvector components assign a numerical value to web pages for their ranking. Finding such a dominant eigenvector on an ever-growing number of web pages becomes a computationally intensive task incompatible with  Moore's Law. We demonstrate that  special-purpose optical machines such as networks of optical parametric oscillators, lasers, and gain-dissipative condensates, may aid in accelerating the  reliable reconstruction of  principal eigenvectors of real-life web graphs. We discuss the feasibility of simulating the PageRank algorithm on  large Google matrices using such unconventional hardware. We   offer alternative rankings based on the minimisation of spin Hamiltonians. Our estimates show that   special-purpose optical machines  may provide dramatic improvements in power consumption over classical computing architectures.
 }
\end{abstract}

\maketitle

\section*{Introduction}

Access to reliable information has always been and will continue to be critical to people's lives and rights. Diverse ways to retrieve information include text, voice, and image queries to search engines, which systematise human knowledge and provide universal access to hundreds of billions of worldwide web pages (or simply web pages) daily \cite{GoogleChannel2020}. At a query time, the most relevant pages are returned in a fraction of a second. Behind such impressive time performance lie significant computational resources that can be divided into two categories.
First, the semantic meaning of a query is analysed by applying traditional information retrieval techniques, combining advances in computer science and statistics, and machine learning methods, including the latest natural language processing algorithms for context analysis \cite{devlin2018bert}. Millions of pages are retrieved with potentially relevant information to the query. Second, before the search happens, the database of publicly available web pages is precomputed and organised by applying hundreds of ranking metrics covering the linking structure, keywords, location, and content freshness of each page. By combining the ranking scores of these two steps, the final order of the most relevant web pages is determined in response to the query \cite{langville2004deeper}.

One ranking algorithm remains in use since the first launch of the Google search engine. The PageRank algorithm \cite{brin1998anatomy,page1999pagerank} evaluates the relative importance of pages by exploiting the web link structure (web graph) solely. The web network is represented as a directed graph, where each page is a node of the graph, and each hyperlink is an edge connecting one page to another. For the entire database of web pages, the PageRank algorithm computes a single score vector, the PageRank vector (or simply PageRank). The algorithm's key underlying assumption is that pages transfer the importance to other pages via links and, hence, the PageRank vector components determine the importance of pages regardless of their textual or visual content and the search query. Mathematically, finding the PageRank vector is equivalent to calculating the principal eigenvector of the link-structure matrix, Google matrix. The general mathematical principles of the PageRank algorithm inspired extensive studies beyond its original use for ranking web search results. A wide range of applications was found in various domains, including social network analysis, recommendation systems, bibliometrics, bioinformatics, DNA sequencing, and distributed computing systems \cite{ermann2015google,gleich2015pagerank}.

The ranking of web pages with the PageRank algorithm, which is connectivity-based and query-independent, does not require real-time processing and is computed in advance. Since the principal eigenvectors can be found in polynomial time, the problem of computing the PageRank vector belongs to the $\mathbb{P}$ complexity class. While being a simple task from the computational complexity theory perspective, the processing of the tens of billions of elements of the link-structure matrix represents a numerical challenge for running the PageRank algorithm on conventional hardware. Considering that the PageRank needs to be updated regularly due to the evolving database with hundreds of new web pages published every second, significant efforts are devoted to exploring efficient ways of computing the PageRank vector and alternative ranking vectors. The principal eigenvector can be found with various linear algebraic techniques \cite{langville2004deeper}, Monte Carlo-type methods running at sub-logarithmic in problem size time \cite{avrachenkov2007monte,bahmani2010fast,sarma2013fast}. A plethora of approaches was suggested for accelerating the original PageRank algorithm based on the power method \cite{kamvar2003extrapolation,kamvar2004adaptive}.

Further development of search technologies requires ever-increasing computational resources. Recent advances were achieved with cloud-based tensor processor unit pods with the power of over 100 petaflops and specialised chips designed to accelerate the training of neural networks. Albeit enough computing resources may be available today, the future demand for prodigious amounts of processing power is beyond traditional hardware. The adiabatic quantum algorithm \cite{garnerone2012adiabatic} and quantum stochastic walks \cite{paparo2012google,tang2020tensorflow} are considered as potential quantum analogues of the PageRank algorithm. Classical physical systems, such as crosspoint resistive memory arrays \cite{sun2020memory}, are proposed for emulating the original PageRank algorithm based on the power method. In another direction of novel computing, various unconventional physical systems are considered as simulators that can minimise spin Hamiltonians. Mapping a real-life optimisation problem into such Hamiltonian and its concomitant minimisation by the natural or guided evolution of the systems promises to solve hard optimisation tasks. The various platforms for such optimisation include optical parametric oscillators \cite{McMahon2016,Inagaki2016}, electronic oscillators \cite{Boehm2019,Chou2019}, memristors \cite{cai2020power}, lasers \cite{Babaeian2019,Pal2019,parto2020realizing,gershenzon2020exact}, photonic simulators \cite{Pierangeli2019,RoquesCarmes2020}, cold atoms \cite{struck2013engineering,anikeeva2020number}, trapped ions \cite{kim2010quantum}, polariton condensates \cite{Berloff2017,kalinin2020polaritonic}, photon condensates \cite{Kassenberg2020}, QED \cite{guo2019sign,marsh2020enhancing}, and others \cite{okawachi2020demonstration,cen2020microwave,dutta2020ising}. While the demonstration of their ability to find the global minima of computationally hard problems faster than the classical von Neumann architecture remains elusive, many of these disparate physical systems can either efficiently perform matrix-vector multiplication \cite{Pierangeli2019,kumar2020large,shen2017deep,prabhu2020accelerating,bernstein2020freely} or mimic the Hopfield neural networks \cite{hopfield1982neural,tait2017neuromorphic,cai2020power}. For a certain choice of parameters, the time evolution of such networks can be viewed as an eigenvalue maximisation problem \cite{Aiyer1990}, which results in finding the energy state dictated by signs of the eigenvector corresponding to the largest eigenvalue of the interaction matrix, i.e. principal eigenvector.

\begin{figure}[t!]
	\centering
	\includegraphics[width=8.6cm]{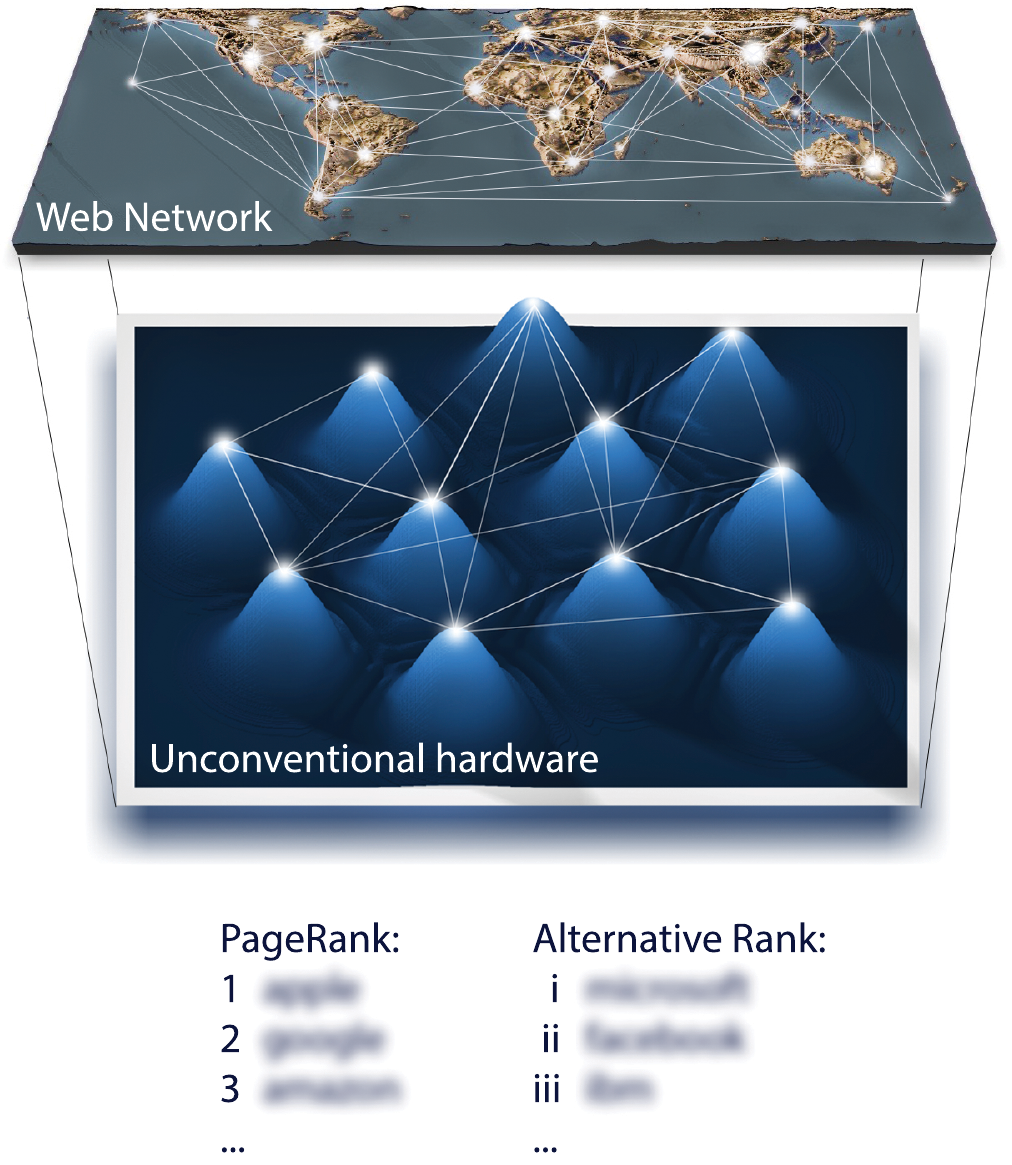}
	\caption{\textbf{The schematics of running ranking algorithms on unconventional hardware.} The link-structure of pages is represented as the Google matrix of the web network (top image). This web network is then mapped to unconventional hardware (middle image) that could be based on a variety of physical platforms, including optical parametric oscillators, lasers, polariton and photon condensates. The time-evolution of unconventional networks can mimic the traditional PageRank algorithm and find the principal eigenvector (PageRank) of the Google matrix or offer alternative rankings based on the minimisation of spin Hamiltonians.
  }
   \label{Fig1}
\end{figure}

This article demonstrates that the Pagerank algorithm can be naturally simulated on unconventional hardware based on a variety of physical systems. We consider networks of optical parametric oscillators, polariton and photon condensates, coupled lasers, as well as the original Hopfield networks and show their ability to efficiently find principal eigenvectors of the Google matrix (see schematics in Fig.~\ref{Fig1}). The system-dependent parameters that ensure the PageRank algorithm's emulation can be found analytically through the stability analysis of dynamical networks and correspond to their operation in the lowest power consumption regime. We confirm networks' ability to reliably find the PageRank vectors by classifying the importance of pages in actual web graphs with sizes from 500 to 3.5 million, including social and university networks. For all datasets, the convergence properties of unconventional networks are shown to be equivalent to the traditional power method. In addition, the unconventional hardware can offer opportunities for exploring alternative rankings. We show that one of such possible rankings could be based on the minimisation of the XY spin Hamiltonian. We further discuss the feasibility of experimental implementations of the large-scale Google matrices on existing unconventional hardware, study the critical noise levels for their reliable operation, and argue about the potential dramatic improvements in power consumption over classical hardware they could bring.

\section*{Ranking Algorithms}

\subsection*{PageRank algorithm}

We start by briefly reviewing the foundations of the PageRank algorithm. The Pagerank algorithm evaluates the importance of web pages based on their connectivity via hyperlinks. The web graph is represented by the Google matrix ${\bf G}$ and the power method is used for finding the PageRank vector ${\bf p}$. In the original algorithm, this method is formulated as \cite{brin1998anatomy}:
\begin{equation}
  {\bf p}^{(k+1)} = {\bf G} \cdot {\bf p}^{(k)}.
  \label{eq:power_method}
\end{equation}
After a certain number of iterations $k$, the power method converges to the principal eigenvector of the matrix ${\bf G}$, which is known as the PageRank vector. The largest components of the principal eigenvector represent the most relevant pages with ranks given by the indices of ordered decreasing components (the PageRank order ${\bf \mathcal{P}}$). For a unique stationary solution of the power method in Eq.~(\ref{eq:power_method}) to exist, the Google matrix ${\bf G}$ is constructed as stochastic and irreducible \cite{langville2004deeper,berkhin2005survey,langville2011google}:
\begin{equation}
  {\bf G} = \alpha {\bf P}^T + {\bf v} [\alpha {\bf d}^T + (1 - \alpha) {\bf e}^T]
  \label{eq:google_matrix}
\end{equation}
where ${\bf P}$ is the transition matrix that represents hyperlink structure of the web, ${\bf d}$ is the dangling vector with ${d_i = 1}$ for zero rows of ${\bf P}$ and $0$ otherwise, ${\bf v}$ represents a personalisation vector, ${\bf e}$ is the unity vector, and $\alpha$ is the teleportation (convergence rate) parameter. For $\alpha \in [0,1)$, the Google matrix has a unique principal eigenvector that corresponds to the largest positive eigenvalue $\lambda_{\rm max} = 1$ \cite{langville2011google,brin2002introduction}. The details of such construction of the Google matrix are outlined in \textit{Materials and Methods}.

To emulate the PageRank algorithm on unconventional hardware, we reformulate the power method of Eq.~(\ref{eq:power_method}) as an iterative scheme on the components $p_i$:
\begin{equation}
  \frac{d p_i}{d t} = - p_i + \sum_{j = 1}^N G_{ij} p_j,
  \label{eq:power_method_iterative}
\end{equation}
whose stationary solution realises the principal eigenvector of the Google matrix. As we shall see below, similar dynamic behaviour is reflected by the operation of many physical systems.

\subsection*{PageRank algorithm on unconventional hardware}

Novel computing paradigms, based on networks of various physical elements from nonlinear oscillators to atoms, may offer a computational advantage over conventional hardware in solving complex optimisation tasks \cite{kalinin2020nonlinear}, many of which can be reformulated as minimisation of discrete or continuous spin Hamiltonians. Unlike the optimisation of hard optimisation problems, calculating the PageRank vector is a polynomially simple task but of large dimensionality. To determine the requirements for unconventional hardware to simulate the PageRank algorithm or, equivalently, to find the principal eigenvector, we formulate the general dynamic network description based on physical systems as:
\begin{equation}
  \frac{d x_i}{dt} = f_i(x_i) + h_i(x_i) \sum_{j = 1}^N \widehat{J}_{ij} g_j(
                \sum_{k = 1}^N{\widetilde{J}_{jk}}x_k),
  \label{eq:general_network}
\end{equation}
where $x_i$ is the real variable that describes a certain measurable physical quantity for each $i$-th element of network, $f_i(\cdot)$ describes the local dynamics, $h_i(\cdot)$ is an amplification function, $g_j(\cdot)$ is an activation function, $\widehat{J}_{ij}$ [$\widetilde{J}_{ij}$] specifies pairwise interactions between $i$-th and $j$-th elements and $\widetilde{J}_{ij}$ [$\widehat{J}_{ij}$] is the identity matrix. This generalised class of unconventional networks is reminiscent of two fundamental modelling approaches of neural networks.
When $\widehat{\bf J}$ is the identity matrix, i.e. $\widehat{\bf J} = {\bf I}$, and interactions between elements are governed by $\widetilde{\bf J} = {\bf J}$, the time-evolution of Eq.~(\ref{eq:general_network}) represents  the static neural network with applications in backpropagation
\cite{pineda1987generalization}. In the opposite case of $\widehat{\bf J} = {\bf J}$ and $\widetilde{\bf J} = {\bf I}$, the networks of physical elements can be viewed as local field neural models \cite{xu2004comparative}. The latter also represents the Cohen–Grossberg model \cite{cohen1983absolute} to which the well-known Hopfield neural networks belong \cite{hopfield1982neural}.
The ability of both the static and local field neural networks to find the principal eigenvector of the interaction matrix ${\bf J}$ originates from the assumption that their nonlinear dynamics can be linearised to:
\begin{equation}
  \frac{d x_i}{dt} = \xi x_i + \beta \sum_{j = 1}^N J_{ij} x_j,
  \label{eq:general_network_linear}
\end{equation}
where $\xi$ and $\beta$ are the system-dependent parameters. The steady states of the linearised network coincide with the minima of the Lyapunov function, which can be introduced as:
\begin{equation}
  \frac{dx_i}{dt} = - \frac{\partial \mathcal{L}}{\partial x_i} \quad {\rm where \ } \mathcal{L} = - \frac{\xi}{2} {\bf x}^T {\bf x} - \frac{\beta}{2} {\bf x}^T {\bf J} {\bf x}.
  \label{eq:gradient_descent_lyapunov}
\end{equation}
The Lyapunov function can be written for a general asymmetric matrix ${\bf J}$ with a Lyapunov equation \cite{kitagawa1977algorithm} although here we assume the matrix to be symmetric for simplicity. The emergence of stable states of Eq.~(\ref{eq:general_network_linear}) can be analysed with the Jacobian matrix with its maximum eigenvalue given by
\begin{equation}
  {\bf \mathcal{J}} = \xi {\bf I} + \beta {\bf J}, \quad \lambda_{\rm max}^{({\bf \mathcal{J}})} = \xi + \beta \lambda_{\rm max}^{({\bf J})}.
  \label{eq:meanfield_jacobian}
\end{equation}
The first nonzero stable state occurs when the maximum eigenvalue of the Jacobian is equal to zero, which leads to the critical value $\xi_{\rm crit}$:
\begin{equation}
  \xi_{\rm crit} = - \beta \lambda_{\rm max}^{({\bf J})}.
  \label{eq:critical_temperature}
\end{equation}
For $\xi < \xi_{\rm crit}$, only the trivial solution ${\bf x} = 0$ exists. At the critical point, the evolution of elements in Eq.~(\ref{eq:general_network_linear}), with time rescaled as $t \rightarrow \beta t$,  and the Lyapunov function $\mathcal{L}$ are expressed as:
\begin{eqnarray}
  \frac{d x_i}{d t} &=& - \lambda_{\rm max}^{({\bf J})} x_i + \sum_{j = 1, N} J_{ij} x_j, \label{eq:general_network_linear_crit}\\
  \mathcal{L} &=& \frac{1}{2} \bigg( \lambda_{\rm max}^{({\bf J})} {\bf x}^T {\bf x} - {\bf x}^T {\bf J} {\bf x} \bigg).
\end{eqnarray}
In this regime, the Lyapunov function is nonnegative, and the network converges toward a stable equilibrium corresponding to the zero minimum value
\begin{equation}
  \min_{\xi = \xi_{\rm crit}} \mathcal{L} = 0 \iff {\bf J} {\bf x} = \lambda_{\rm max}^{({\bf J})} {\bf x}.
\end{equation}
In case of the Google matrix, the largest eigenvalue is equal to one, namely $\lambda_{\rm max}^{({\bf J})} = \lambda_{\rm max}^{({\bf G})} = 1$, and the PageRank vector is represented by the network amplitudes at the steady state of equations
\begin{equation}
  \frac{d x_i}{dt} = - x_i + \sum_{j = 1}^N G_{ij} x_j,
  \label{eq:pagerank_linear}
\end{equation}
that is equivalent to the iterative scheme of the power method in Eq.~(\ref{eq:power_method_iterative}). Starting with any initial conditions, the solution of Eq.~(\ref{eq:pagerank_linear}) will always converge to the equilibrium point of the system corresponding to the principal eigenvector of the Google matrix.

As a demonstration of calculating the PageRank vector on unconventional hardware, we consider networks of elements based on several physical systems, including optical parametric oscillators (OPOs), lasers, polariton and photon condensates. These gain-dissipative systems achieve coherent states when the gain exceeds the losses. At the coherence threshold, when $\xi = \xi_{\rm crit}$, the principal eigenvector of the Google matrix can be reconstructed from the network amplitudes of the first stable nonzero steady state. Besides, the PageRank algorithm can be emulated using the Hopfield networks, which can be implemented with photonic \cite{tait2017neuromorphic} and electronic \cite{cai2020power} systems. For showing a possible robust emulation of the PageRank algorithm on unconventional hardware, the system-dependent parameter configurations are found analytically for all considered networks whose dynamics at the lowest power consumption regime is equivalent of Eq.~(\ref{eq:pagerank_linear}), as discussed in \textit{Materials and Methods}.

\subsection*{Alternative ranking algorithms on unconventional hardware}
Various physical systems are widely studied as unconventional computing platforms for solving hard optimisation problems. To find optimal solutions of hard problems, the densities of all network elements have to be equilibrated at the coherence threshold \cite{leleu2017combinatorial,kalinin2018global}, which requires an implementation of the active feedback scheme. Without this feedback mechanism, certain classes of problems can still be globally optimised. For example, global minima can be associated with the eigenvector corresponding to the largest eigenvalue of the coupling matrix for some problems that are also considered trivial and can be solved in polynomial time \cite{kalinin2020complexity}. As we argue in this paper, the density inhomogeneity at the coherence threshold could be used for reconstructing the PageRank vector on unconventional hardware. Also, the presence of the feedback mechanism could open opportunities for designing alternative rankings based on spin Hamiltonian minimisation. As an example, we consider an alternative ranking, XYRank, based on the minimisation of the XY Hamiltonian, that can be achieved in various physical systems including the lasers \cite{Babaeian2019,Pal2019,parto2020realizing,gershenzon2020exact}, polariton \cite{Berloff2017,kalinin2020polaritonic} and photon \cite{Kassenberg2020} condensates. The XY Hamiltonian is formulated as:
\begin{equation}
  H_{\rm XY} = - \sum_{ij} J_{ij} \cos(\theta_i - \theta_j),
\end{equation}
where $\theta_i \in [0, 2\pi)$ is the phase of the $i$-th element. The minimisation of the XY model can be realised with the gain-dissipative networks \cite{kalinin2018global}:
\begin{eqnarray}
  \frac{d \psi_i}{dt} &=& (\gamma_i - |\psi_i|^2)\psi_i + \sum_{j = 1}^N J_{ij} \psi_j, \label{eq:gd_xy_model_psi}\\
  \frac{d \gamma_i}{dt} &=& \epsilon (\rho_{\rm th} - |\psi_i|^2)
  \label{eq:gd_xy_model_gamma}
\end{eqnarray}
where $\psi_i$ represents the complex amplitude of the $i$-th element, $\gamma_i$ is the effective gain rate (includes linear losses), $J_{ij}$ are the pairwise interactions between the $i$-th and $j$-th elements, $\epsilon$ is the rate for gain adjustments, and  $\rho_{\rm th}$ is the a priori set occupation of each element. The second equation ensures that densities of all elements are equilibrated at the steady-state and, hence, the minima of the XY Hamiltonian are realised with phases recovered from the $\psi_i = \sqrt{\rho_i} \exp[i\theta_i]$). In the case of the Google matrix ${\bf J} = {\bf G}$, the alternative ranking is based on the individual gains $\gamma_i$, which can take negative and positive values. The lower is the gain $\gamma_i$, the higher importance is assigned to the $i$-th element. We call this alternative ranking the XYRank since the minimum of the total power gain $\sum_i \gamma_i$ corresponds to the minimum of the XY Hamiltonian \cite{kalinin2018networks}.

\section*{Numerical Simulations of Unconventional Networks}

\textbf{Emulating PageRank algorithm with unconventional networks.}
To demonstrate a reliable operation of unconventional hardware for computing the PageRank vectors, we simulate the Hopfield neural networks and networks of OPOs, lasers, polariton and photon condensates across various datasets. We consider the Google matrices from small size $N = 500$ to larger sizes up to $N = 3.5$ million based on real web graphs of universities, social Networks, frequently co-purchased products, the Wikipedia top categories, and others (see \textit{Materials and Methods} for a full description of databases). For all datasets, the ranking vectors obtained using unconventional networks are compared with the PageRank vectors computed using the original PageRank algorithm based on the power method. We use a standard metric to measure the correspondence between two rankings, namely the Kendall rank correlation coefficient (Kendall's tau) \cite{kendall1945treatment}:
\begin{equation}
  {\rm Kendall's \  tau} = \frac{C - D}{\sqrt{(C + D + T_1) (C + D + T_2)}} \in [-1, \ 1],
  \label{eq:kendall_tau}
  \nonumber
\end{equation}
where $C$ is the number of concordant pairs, $D$ is the number of discordant pairs, $T_k$ is the number of ties only in the $k$-th ranking. Concordant and discordant pairs describe the relationship between pairs of elements from two rankings: the pair $(i,j)$ is concordant if both methods rate the $i$-th element higher than the $j$-th. A tie occurs for the pair $(i,j)$ when a method assigns equal weights to both elements, while the pairs of elements with equal weights in both rankings do not contribute to either number of ties $T_k$. The larger positive values of Kendall's tau correspond to the stronger agreement between rankings, larger negative values indicate the reverse order of rankings, and near-zero values reflect no correlation between rankings.
\begin{figure}[t!]
	\centering
	\includegraphics[width=8.6cm]{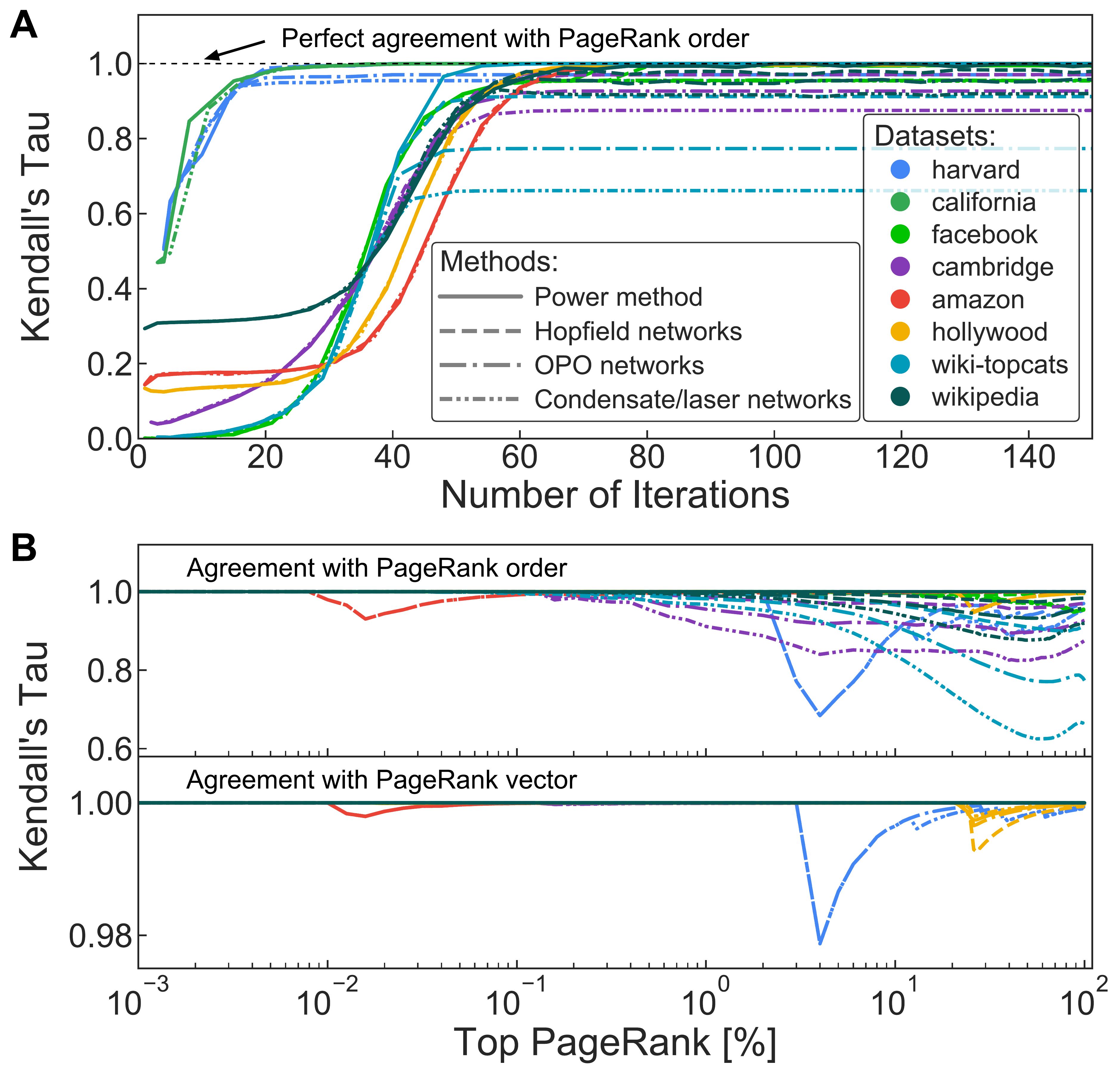}
	\caption{\textbf{Emulation of the PageRank algorithm with unconventional networks}. (A) Kendall's tau, as a measure of the agreement between the PageRank order computed with the power method and the rankings obtained using unconventional networks, is shown as a function of the number of iterations for a variety of web graphs. The sizes of graphs vary from $N = 500$ for the university network (`harvard') to $N=3.5$ million for the Wikipedia database (`wikipedia'). The unconventional networks are represented by Hopfield networks and networks of optical parametric oscillators (OPOs), condensates (polariton and photon), and lasers. (B) Kendall's tau distribution is shown as a function of the highest-ranked elements for the PageRank order (top) and the PageRank vector (bottom). The teleportation parameter is fixed across all datasets to $\alpha = 0.85$.
  }
   \label{Fig2}
\end{figure}

The Kendall's tau between the PageRank order, which corresponds to the indices of the sorted components of the principal eigenvector computed using the power method, and ranks obtained with unconventional networks, is shown in Fig.~\ref{Fig2}A as a function of the number of iterations of all methods. The noticeable discrepancies between rankings of several methods, e.g. networks of OPOs and condensates (lasers), on several datasets, e.g. `cambridge' and `wiki-topcats', originate from the elements with the lowest importance as reflected in Fig.~\ref{Fig2}B(top). Computing Kendall's tau between the PageRank vector and amplitude distributions of the steady states in unconventional networks leads to the occasional minor deviations from the perfect agreement with the PageRank algorithm, as shown in Fig.~\ref{Fig2}B(bottom). Despite nonlinearities, the principal eigenvectors of the Google matrices are reconstructed with high accuracy by all networks.

The varying agreement between methods for the PageRank order ($\mathcal{P}_i$) and PageRank vector ($p_i$) is caused by the processing of ties in Kendall's tau calculation. In the latter case, different rankings can have multiple ties for the same pairs of elements that do not contribute to Kendall's tau, while redistributions of these ties into the concordant and discordant pairs decrease the correlation between rankings in the former case. Hence, we conclude that all unconventional networks successfully produce orderings in strong agreement with the PageRank algorithm.

\textbf{Convergence properties of unconventional networks.}
Given the size of web graphs in tens of billions, the rate of convergence, i.e. the number of iterations required to find the solution, is a crucial quantity to optimise. The asymptotic rate of convergence of the power method in the Pagerank algorithm is governed by the subdominant eigenvalue of the Google matrix, which is equal to $\alpha$ \cite{haveliwala2003second}. Consequently, the proper choice of parameter $\alpha$ for the Google matrix ensures the exponentially fast convergence of the power method given by  Eq.~(\ref{eq:power_method}) due to the finite gap $(1 - \alpha)$ between the largest and the second-largest eigenvalues. For web graphs, the common eigenvalue gap belongs to the interval $\alpha \in [0.85, 0.9]$
\cite{page1999pagerank,kamvar2003extrapolation,haveliwala2003second}, while for other applications optimal values can be lower. For instance, $\alpha = 0.3$ is reported for genes ranking \cite{winter2012google}. In addition to the convergence properties, the choice of different values of $\alpha$ places a different emphasis on the connectivity structure and can result in vastly different principal eigenvectors.

\begin{figure}[b!]
 \centering
 \includegraphics[width=8.6cm]{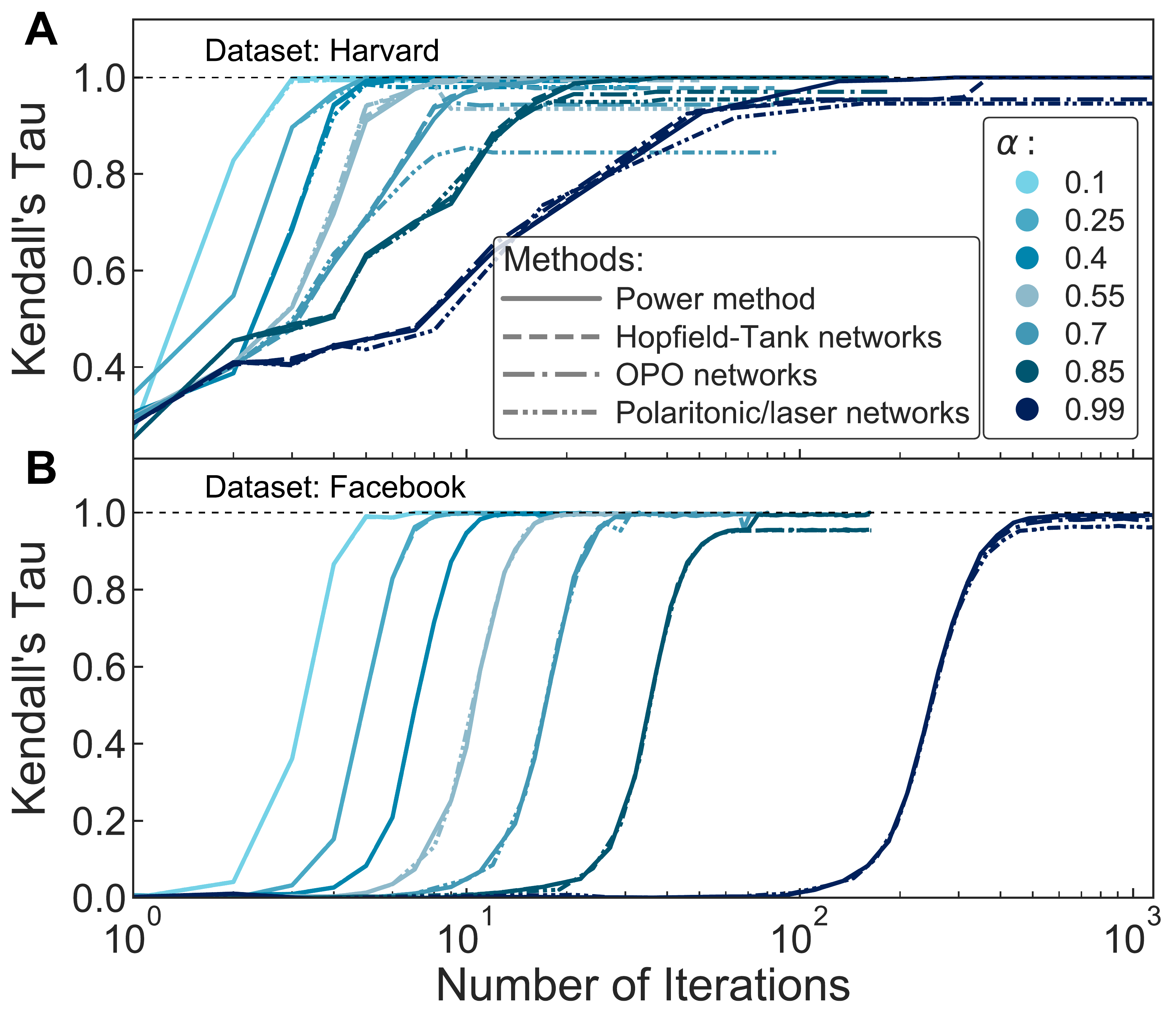}
 \caption{\textbf{Convergence rate for the PageRank algorithm with unconventional networks}. The Kendall's tau dependence is shown as a function of the number of iterations for unconventional networks at different teleportation parameter values $\alpha$ ranging from $0.1$ (light blue) to $0.99$ (dark blue) for two datasets: the Harvard University network (A) and the Facebook social network (B). The dynamics of networks is based on the Hopfield-Tank algorithm, optical parametric oscillators (OPOs), polariton condensates and lasers. The unit Kendall's tau level is taken with respect to the power method.
  }
   \label{Fig3}
\end{figure}

To cover various possible applications and to demonstrate the robustness of physical systems in finding principle eigenvectors, we analyse the convergence rate of unconventional networks at different values $\alpha$ in Fig.~\ref{Fig3}. For both representative datasets, Kendall's tau between PageRank orders and ranks obtained in unconventional networks
shows similar to the traditional power method dependence on the number of iterations across all considered networks. Computing Kendall's tau for the PageRank vector leads to the perfect convergence agreement with unconventional networks (see \textit{Materials and Methods} for details).

\textbf{Convergence properties of unconventional networks in the presence of noise.}
For the PageRank algorithm based on the power method, the initial state is usually chosen as a uniform distribution ensuring the fast and deterministic convergence to the principal eigenvector. Although the same homogeneous initial state is used for all unconventional networks in Fig.~\ref{Fig2} and Fig.~\ref{Fig3} for a transparent comparison,
such an assumption is unlikely to hold for most physical systems operating in noisy environments. Starting with random initial conditions taken from the Gaussian distribution $\mathcal{N}(0, 1/N)$, the number of iterations required for convergence to a given precision increases by about 5-35 iterations for all considered methods and datasets. With the additional presence of the intrinsic noise during the time-evolution of network elements, the lowest components of the principal eigenvector are reordered first for small noise amplitudes, with higher rank components being affected for larger noises.

For example, in the case of the web graph of co-purchased products (\textit{`amazon'}) with size $N = 400727$, the small noise at the level of the lowest-ranked elements leads to the strong agreement for the 100 highest ranked elements with Kendall's tau of $0.99$ and overall ranking coefficient of $0.79$. For the noise at the median level of the principal eigenvector components, the good correspondence is still achieved for the top 100 elements with Kendall's tau of $0.94$ whereas the total Kendall's tau drops to $0.34$. Since the PageRank vector values can vary by several orders of magnitude and only the correct order matters, one may expect that even in networks based on noisy physical systems, the highest-rated elements could be found with high precision.

\textbf{Alternative ranking: minimisation of the XY Hamiltonian.}
Physical systems can provide unconventional hardware to mimic the original \mbox{PageRank} algorithm and be used to explore alternative rankings. One of such new rankings could be based on the minimisation of the XY Hamiltonian, i.e. XYRank. We show the relation between the traditional \mbox{PageRank} and XYRank distributions in Table~\ref{table:pr_top20}. The highest-ranked elements are simply reshuffled for web graphs \textit{`harvard'}, \textit{`facebook'}, and  \textit{`wiki-topcats'}. In the case of the \textit{`california'} dataset, several of the PageRank positions are given much lower importance with respect to the XYRank, while the top XYRank positions still belong to the highly-rated pages of the PageRank distribution (see \textit{Materials and Methods}). Understanding which ranking algorithm is best and whether XYRank can lead to better search results requires a detailed ranking analysis beyond algorithmic methods. In commercial search engines, thousands of trained external raters evaluate search quality results on various datasets and queries, even for small changes in ranking algorithms. The ability to have a platform that emulates the traditional PageRank algorithm and offers alternative rankings could allow unconventional hardware to safely replace traditional computing architectures and facilitate the development of new search algorithms.
\begin{table*}[t]
  \begin{ruledtabular}
    \caption{The highest 10 PageRank positions are shown for datasets \textit{`harvard'}, \textit{`california'}, \textit{`facebook'}, and \textit{`wiki-topcats'} of size $N = 500$, $N = 9664$, $N = 22470$, and $N = 1791489$. The identical PageRank distributions are found between the original PageRank algorithm based on the power method and simulations of Hopfield networks and networks based on optical parametric oscillators, lasers, polariton and photon condensates. The alternative ranking (XYRank) is computed by minimising the XY Hamiltonian. The difference between the two ratings is indicated by green (red) arrows showing the shift in the XYRank towards a higher (lower) rating by a certain number of positions with respect to the PageRank.}
    \centering
    \label{table:pr_top20}
    \begin{tabular}{c l l l l}
      PageRank & Harvard & XYRank & California & XYRank \\
      \hline
      1 & www.harvard.edu & $-$ & www.ucdavis.edu/ & $\color{red}\Downarrow58$ \\
			2 & www.hbs.edu & $\color{red}\Downarrow6$ & search.ucdavis.edu/ & $\color{red}\Downarrow421$ \\
			3 & search.harvard.edu:8765/custom/.. & $\color{green}\Uparrow1$ & www.california.edu/ & $\color{red}\Downarrow22$ \\
			4 & www.med.harvard.edu & $\color{green}\Uparrow1$ & home.netscape.com/comprod/mirror/.. & $\color{green}\Uparrow2$ \\
			5 & www.gse.harvard.edu & $-$ & www.berkeley.edu & $-$ \\
			6 & www.hms.harvard.edu & $\color{red}\Downarrow3$ & www.linkexchange.com/ & $\color{green}\Uparrow5$ \\
			7 & www.ksg.harvard.edu & $\color{green}\Uparrow1$ & www.berkeley.edu/ & $\color{green}\Uparrow3$ \\
			8 & www.hsph.harvard.edu & $\color{green}\Uparrow4$ & www.uci.edu/ & $\color{red}\Downarrow173$ \\
			9 & www.gocrimson.com & $\color{red}\Downarrow5$ & www.ca.gov & $\color{red}\Downarrow7$ \\
			10 & www.hsdm.med.harvard.edu & $\color{red}\Downarrow12$ & www.lib.uci.edu/ & $\color{red}\Downarrow291$ \\
      \hline
      PageRank & Facebook & XYRank & Wikipedia (top categories) & XYRank \\
      \hline
      1 & Facebook & $\color{red}\Downarrow2$ & United States & $-$ \\
      2 & Sir Peter Bottomley MP & $\color{green}\Uparrow1$ & France & $-$ \\
      3 & The White House & $\color{red}\Downarrow9$ & United Kingdom & $\color{red}\Downarrow8$ \\
      4 & The Obama White House & $\color{red}\Downarrow11$ & Canada & $\color{red}\Downarrow1$ \\
      5 & U.S. Army & $\color{red}\Downarrow6$ & Germany & $\color{red}\Downarrow5$ \\
      6 & U.S. Army Chaplain Corps & $\color{green}\Uparrow1$ & World War II & $\color{red}\Downarrow18$ \\
      7 & Joachim Herrmann & $\color{green}\Uparrow5$ & English language & $\color{red}\Downarrow16$ \\
      8 & Barack Obama & $\color{red}\Downarrow1$ & Australia & $\color{red}\Downarrow5$ \\
      9 & European Parliament & $\color{red}\Downarrow5$ & Italy & $\color{red}\Downarrow9$ \\
      10 & Manfred Weber & $\color{green}\Uparrow4$ & India & $\color{red}\Downarrow10$ \\
    \end{tabular}
  \end{ruledtabular}
\end{table*}

\begin{figure}[b!]
 \centering
 \includegraphics[width=8.6cm]{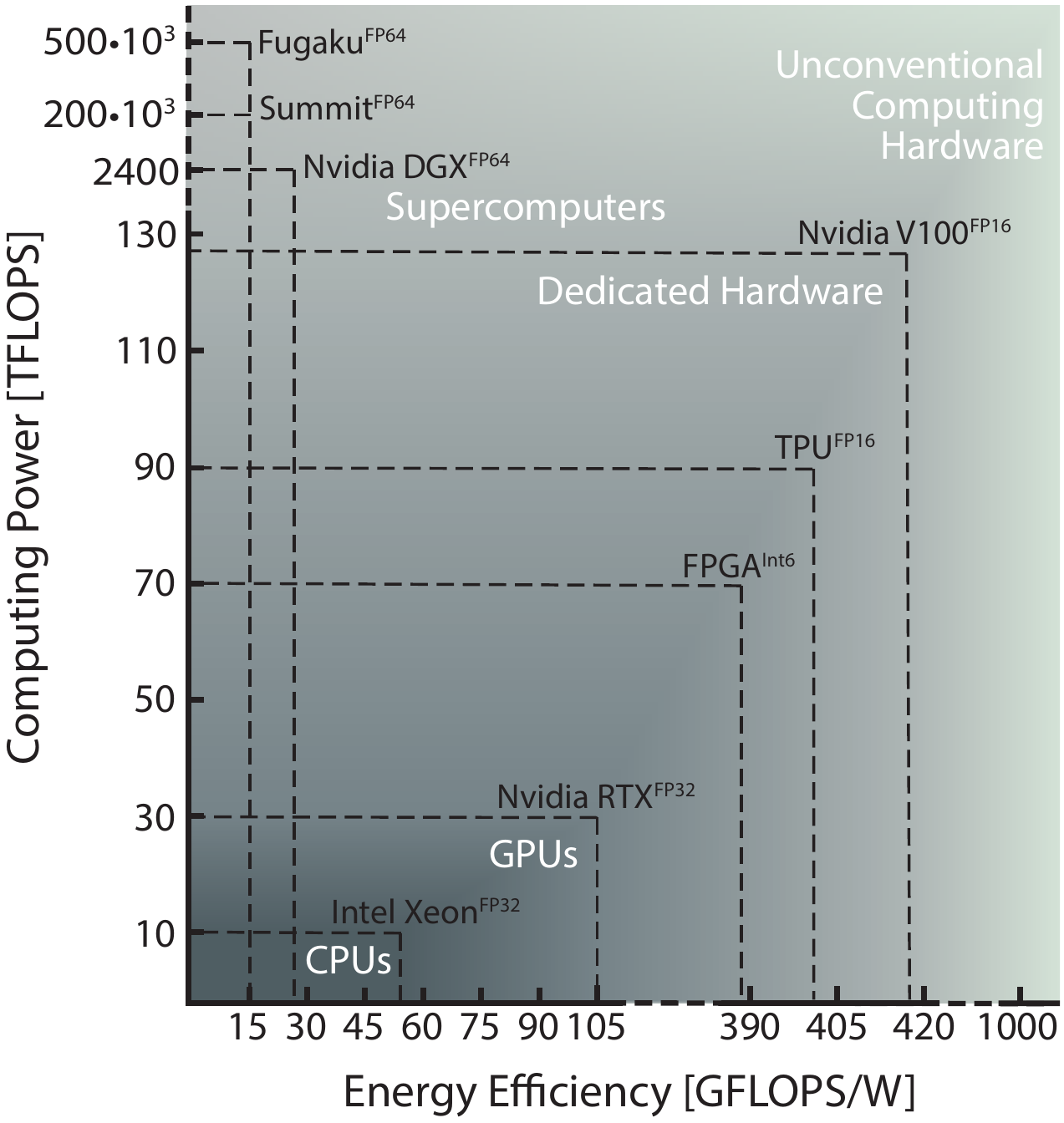}
 \caption{\textbf{Computing power and energy efficiency of computing hardware}. The schematic distribution of the processing power as a function of energy efficiency is shown for several conventional computing devices, including CPUs, GPUs, FPGAs, and supercomputers. Unconventional computing devices based on optical systems could provide orders of magnitude improvements in time and energy performance.
  }
   \label{Fig4}
\end{figure}

\textbf{Computing power and energy efficiency.}
The evolving nature of the Internet requires regular updates of PageRank distributions. Whereas 20 years ago, almost half of all web pages were updated at weekly intervals \cite{cho1999evolution}, nowadays constant changes in the web structure can occur within an hour or even a minute. As an estimate, the regular updates of the PageRank vector on a minute scale for a 10 billion size matrix would result in the average annual electricity consumption of about $3.5 \cdot 10^5 \ {\rm kWh}$ on dedicated hardware, such as the tensor processing units (TPUs), see Fig.~\ref{Fig4}, that is equivalent to charging up about 300 electric cars for one year (see \textit{Materials and methods} for details). In addition to updating the global web network, the monetary success of many other platforms depends on how often similar to the PageRank ratings are calculated. For example, SalesRank needs to be updated hourly to reflect the purchase history of Amazon products.

To keep up with the dynamic changes of the web structure and its growing size, unconventional hardware can offer a time and energy-efficient platform for performing such intensive computations. The unconventional hardware typically benefits from inherent computational parallelism and possible quantum speedup. The typical improvements to the power method, such as reduced recalculation of converged pages \cite{kamvar2004adaptive}, are naturally embedded in physical systems. In general, the time performance and energy consumption of computing \mbox{PageRank} and alternative ranks on unconventional hardware depend on the type of possible architectures.

For a hybrid (or active) coupling scheme, when interactions in the network are created using traditional devices, the performance of unconventional hardware is dictated by characteristics of the classical counterpart. For example, arbitrary coupling configurations can be implemented in OPO-based networks by matrix multiplication on a field-programmable gate array (FPGA) \cite{McMahon2016}, which limits the time and energy performance of the optical system to the operational characteristics of FPGA. For polariton networks \cite{kalinin2020polaritonic} and other photonic systems \cite{Pierangeli2019}, couplings could be realised using a spatial light modulator, which also restricts their time and energy performance.

To take full advantage of the capabilities of physical systems underlying unconventional hardware, pure optical architectures with passive schemes for creating interactions could be considered. In such all-optical passive networks, couplings do not require reconfiguration at each iteration, and the performance is determined solely by the characteristics of physical systems. While such coupling schemes are more difficult to engineer, there exist proposals of all-optical OPO \cite{Inagaki2016} and polaritonic machines \cite{kalinin2020polaritonic}, where the latter offers a platform with picosecond runtime. The programmable photonic processors provide passive integrated circuits with subnanosecond operation time scale \cite{RoquesCarmes2020}.

The presence of inherent nonlinearities in some physical systems requires operation near the coherence threshold to simulate the PageRank algorithm with high accuracy. Nonlinearities are introduced intentionally in other physical computing platforms and can be omitted to model the PageRank algorithm. For example, the \mbox{PageRank} can be calculated by performing optical matrix multiplications that support beyond ${\rm GHz}$ clock rates \cite{shen2017deep,prabhu2020accelerating,bernstein2020freely}.

To estimate the energy efficiency of unconventional hardware, we note that the power intensity required for creating an element of a network is on the order of milliwatts for most optical systems \cite{prabhu2020accelerating}. For example, a milliwatt laser power is usually required for exciting one micron-sized polariton condensate. Hence, the approximate power scaling with network size could be taken as ${P \sim N \ {\rm mW}}$. The computational complexity of the PageRank algorithm is governed by the matrix-vector multiplication product in the power method and can be expressed as $\mathcal{O}(mnN)$, where $m$ is the average connectivity of the web graph, and $n$ is the number of iterations required for convergence. Then the number of floating-point operations per second (FLOPS) for unconventional hardware operating at picosecond-nanosecond time scale could be in the range of ${m n N \cdot [10^{9}, \ 10^{12}] \ {\rm FLOPS}} $ with the energy efficiency of
\begin{equation}
  {\rm Energy \ efficiency} \approx mn \cdot [1, \ 10^3]  \ \frac{\rm TFLOPS}{\rm W}.
\end{equation}
Evidently, unconventional hardware with passive coupling schemes could provide orders of magnitude more energy-efficient performance than conventional computing architectures, see Fig.~\ref{Fig4} for their power and efficiency comparison.

\section*{Discussion}

Information has the power to unlock unseen opportunities for people. Nevertheless, the dynamic nature and volume of information available on the Internet make the search challenging to improve. For web graphs with billions of nodes, even one computation of the PageRank vector is a computationally intensive task. The goal of realising real-time personalised and topic-sensitive rankings, when the PageRank is calculated for the nonuniform teleportation vector, requires the fast generation of a large set of \mbox{PageRank} vectors simultaneously, which is practically infeasible with traditional computing hardware. Unconventional hardware could enhance both computational speed and energy efficiency of PageRank calculation over the state-of-the-art classical computing architectures, thereby facilitating the exploration of new approaches to personalised search engine schemes. As an example of non-traditional platforms, we demonstrate that networks of elements based on various physical systems, including optical parametric oscillators, polariton and photon condensates, lasers, and other systems that mimic the behaviour of the Hopfield neural networks, can be used to simulate the PageRank algorithm on large-size web graphs reliably. These platforms can also provide opportunities for studying alternative rankings based on the minimisation of spin Hamiltonians, opening up unexplored routes for improving search algorithms.

Analogue optical platforms may become a new class of multi-purpose computing architectures with ultra-low power consumption in the near future. Besides grand applications of solving practically relevant hard optimisation problems, training and running neural networks, this non-traditional hardware could be exploited to accelerate existing search engine techniques and explore novel ranking approaches. Consequently, unconventional hardware based on physical systems may represent an efficient computing and long-term sustainable paradigm for continued innovation in search and other applications.

\section*{Materials and Methods}

\subsection*{1. Google Matrix Construction Details}

The stages for constructing the Google matrix ${\bf G}$:
\begin{eqnarray}
  {\bf G} = ({\bf P}^{''})^T &=& [\alpha {\bf P} + (\alpha {\bf d} + (1 - \alpha) {\bf e}) {\bf v}^T]^T \nonumber \\
  &=& \alpha {\bf P}^T + {\bf v} [\alpha {\bf d}^T + (1 - \alpha) {\bf e}^T]
\end{eqnarray}
could be divided into the following steps (see \cite{page1999pagerank,haveliwala2003second,kamvar2003extrapolation,langville2011google} for more details):
\begin{itemize}
  \item[i.] ${\bf P}$ is the directed (undirected in rare cases) transition matrix, whose nodes represent web pages and the directed edges correspond to hyperlinks, with elements expressed as $P_{ij} = A_{ij} / deg(i)$ when $deg(i) > 0$ and $P_{ij}  = 0$ otherwise. Here ${\bf A}$ is the adjacency matrix with $A_{ij} = 1$ when there is a link from page $i$ to page $j$, and $A_{ij} = 0$ otherwise, and $deg(i)= \sum_j A_{ij}$ is the number of outgoing links of a page $i$ (out-degree). Thanks to such normalisation, the matrix elements $P_{ij}$ represent probabilities of moving from page $i$ to page $j$ in one time-step.
  \item[ii.] The stochastic matrix ${\bf P}^{'}$ is constructed from the transition matrix ${\bf P}$ as ${\bf P}^{'} = {\bf P} + {\bf d} {\bf v}^T$, where ${\bf d}$ is the dangling vector with $d_i = 1$ for zero rows of ${\bf P}$ and $0$ otherwise. The dangling nodes are the nodes without outlinks in the transition matrix. Such nodes commonly occur in practice and can be attributed, for example, to the unexplored (`crawled') web pages that are added to the web graph. The uniform vector ${\bf v}$ adds artificial links by connecting uniformly dangling pages to all pages in the web graph, while the non-uniform choice of ${\bf v}$ represents a personalisation vector. For such nonnegative row-stochastic matrix ${\bf P}^{'}$, the principal eigenvector corresponds to the eigenvalue $\lambda = 1$ which could be degenerate, while such degeneracy can prevent the convergence of the power method.
  \item[iii.] The stochastic irreducible matrix ${\bf P}^{''}$ is formed as ${\bf P}^{''} = \alpha {\bf P}^{'} + (1 - \alpha) {\bf e} {\bf v}^T$, where ${\bf e}$ is the unity vector. The matrix is irreducible (strongly connected) since every page is now directly connected to every other page. The irreducibility adjustment also ensures that ${\bf P}^{''}$ is primitive, which guarantees the existence of the unique stationary PageRank vector for $\alpha < 1$, to which the power method will converge regardless of the initial distribution. The parameter $\alpha$ is known as the damping (teleportation) factor since for a non-dangling page $i$ a random web surfer can not only follow one of the available outlinks with the probability of $\alpha$ but can also jump (teleport) to any other page $j$ with the probability of $(1 - \alpha) v_j$. Hence, the probability vector ${\bf v} > 0$ is known as the teleportation vector when chosen to be uniform or personalisation vector otherwise.
\end{itemize}

For computational efficiency, the Google matrix ${\bf G}$ is never explicitly formed. Instead, the PageRank vector can be calculated with the power method as \cite{kamvar2003extrapolation}:
\begin{equation}
  {\bf p}^{(k+1)} = {\bf G} \cdot {\bf p}^{(k)} = \alpha {\bf P}^T \cdot {\bf p}^{(k)} + {\bf v} [\alpha {\bf d}^T {\bf p}^{(k)} + (1 - \alpha)],
  \label{eq:power_method_sparse}
\end{equation}
where the normalisation condition ${\bf e}^T {\bf p}^{(k)} = 1$ is assumed and the advantage of the sparse matrix-vector multiplication could be taken of. The PageRank algorithm is also directly linked to the mathematical properties of Markov chains and Perron-Frobenius operators \cite{brin2002introduction,langville2011google} and can be viewed as a stationary probability distribution for the Markov chain induced by a random walk on the web graph.

\subsection*{2. Feasibility and requirements of embedding the Google matrix in physical systems}
To discuss the possibility to embed the Google matrix on unconventional hardware, we comment on all three stages of construction of the Google matrix.

The first step requires an ability to realise a sparse adjacency matrix in a physical platform. It should be possible to create directed interactions between arbitrary nodes in the network with about 10 to 100 connections per node. This requirement is feasible for most of the physical systems considered in this work since the couplings can be usually organised by external means. For example, the field-programmable gate arrays are used for OPOs, and spatial light modulators can be used for polariton condensates. Such number of interactions could be possibly harder to realise in passive coupling schemes.

In the second step, some nodes (dangling nodes) should be equally connected to all other nodes with a small interaction strength. Given the possibly large number of the dangling nodes, this could be an expensive step to complete with purely optical means from the engineering perspective, although it is straightforward with externally supplied couplings. Alternative ways to create a stochastic matrix may need to be explored to implement such interactions in physical systems efficiently.

In the third step, the irreducible matrix is created. Creating such a strongly-connected matrix may look even more challenging to engineer than a few fully-connected nodes in the second step. Fortunately, there could be a simple way to do this. The method of minimal irreducibility \cite{tomlin2003new} was proven to be equivalent both in theory and in computational efficiency to the maximally irreducible method (that is used used in the third step) \cite{langville2004deeper}. An additional node is added to the network and connected to all other nodes in this alternative method. Such an extra node plays the role of a teleportation state: there is a small probability of transitioning to and out of this state. Unlike maximal irreducibility, the minimal irreducibility could be naturally realised in many physical systems by applying a bias, for instance, a small uniform magnetic field to all network elements. Also, a nonuniform magnetic field could allow one to emulate the personalised PageRank algorithm.


\subsection*{3. The PageRank algorithm on unconventional hardware}

\textbf{Networks of optical parametric oscillators.}

Network of coupled optical parametric oscillators (OPOs) represents an unconventional gain-dissipative platform \cite{wang2013coherent,marandi2014network} whose simplified dynamics is governed by the equations
\begin{equation}
  \frac{dx_i}{dt} = - x_i^3 + (p - 1) x_i + \beta \sum_{j = 1}^N J_{ij} x_j,
  \label{eq:opo_networks}
\end{equation}
where $p$ represents the linear gain, $J_{ij}$ are the interactions between oscillators, the linear and nonlinear losses are normalised. The OPO-based simulator is proposed initially as a coherent Ising machine since two possible phase states exist for each nonlinear oscillator above a certain pumping threshold, and, hence, these states can be interpreted as binary spins. For degenerate optical parametric oscillators (DOPOs) in a fibre \cite{McMahon2016}, arbitrary coupling connections between any two spins can be realised in the feedback loop on a field-programmable gate array (FPGA).

For mimicking the PageRank algorithm,  networks of OPOs should be operating at the lowest loss regime. In the case of the Google matrix, this critical regime  corresponds to \cite{wang2013coherent,leleu2017combinatorial}
\begin{equation}
  p_{\rm crit} - 1 = - \beta \lambda_{\rm max}^{({\bf G})} = - \beta.
\end{equation}
The linear stability analysis of Eq.~(\ref{eq:opo_networks}) is similar to that of Eq.~(\ref{eq:general_network_linear}) at
$\xi_{\rm crit} = p_{\rm crit} - 1$. Hence, the PageRank vector is realised by the OPOs amplitudes $x_i$ in the steady state of equation:
\begin{equation}
  \frac{dx_i}{dt} = - \frac{1}{\beta} x_i^3 - x_i + \sum_{j = 1}^N G_{ij} x_j.
  \label{eq:opo_networks_pagerank}
\end{equation}
To get this equation, the time is rescaled as $t \rightarrow \beta t$ in Eq.~(\ref{eq:opo_networks}).

\textbf{Networks of polariton and photon condensates.}

Lattices of polariton condensates are another gain-dissipative unconventional hardware that has recently been proposed for the minimisation of discrete and continuous spin Hamiltonians \cite{Berloff2017,kalinin2020polaritonic}. Exciton-polaritons, or simply polaritons, are quasi-particles that arise from the superposition of photons and excitons in semiconductor microcavities. Despite being a nonequilibrium system, polaritons are bosons that obey Bose-Einstein statistics and form a coherent macroscopic state above critical pumping \cite{kasprzak2006bose}. The polariton mass depends on the microcavity structure. The condensates can be realised not only at cryogenic temperatures but also at room temperatures in organic structures \cite{cookson2017yellow}. The physics of polariton condensates resembles another unconventional computing system based on photon condensates confined in a dye-filled optical microcavity \cite{klaers2010bose,Kassenberg2020}. These networks of gain-dissipative condensates can be realised in experiments using a spatial light modulator \cite{wertz2010spontaneous} with many techniques proposed and engineered for controlling couplings between condensates \cite{schneider2016exciton,dung2017variable,kalinin2020toward,alyatkin2020optical}.

The time-evolution of gain-dissipative condensates is derived from the space and time-resolved mean-field equations \cite{kalinin2018networks,kalinin2019polaritonic}
and described by the Stuart-Landau equations:
\begin{equation}
  \frac{d \psi_i}{dt} = -i U |\psi_i|^2 \psi_i + (\gamma - |\psi_i|^2)\psi_i + \sum_{j = 1}^N J_{ij} \psi_j,
  \label{eq:gd_condensates}
\end{equation}
where $\psi_i$ represents the complex amplitude of the $i$-th condensate, $U$ stands for the strength of nonlinear interactions, $\gamma$ is the effective pumping rate (includes linear losses), and $J_{ij}$ are pairwise interactions between condensates. Generally, the interactions are complex-valued, although the imaginary part is relatively small and, hence, can be neglected.

For simulating the PageRank algorithm with networks of gain-dissipative condensates, we show the equivalence of stability of linearised equations Eq.~(\ref{eq:gd_condensates}) to Eq.~(\ref{eq:general_network_linear}). By substituting $\psi_i = \sqrt{\rho_i} \exp[i\theta_i]$ and separating real and imaginary parts in Eq.~(\ref{eq:gd_condensates}), we obtain
\begin{equation}
  \begin{aligned}
       {\rm Re} & \quad \frac{d \sqrt{\rho_i}} {dt} = (\gamma - \rho_i) \sqrt{\rho_i} + \sum_{j = 1}^N J_{ij} \sqrt{\rho_j} \cos(\theta_j - \theta_i), \\
       {\rm Im}  &\quad \frac{d \theta_i} {dt} = - U \rho_i + \sum_{j=1}^N J_{ij} \sqrt{\frac{\rho_j} {\rho_i}} \sin(\theta_j - \theta_i),
       \label{eq:gd_cond_real_and_imag}
  \end{aligned}
\end{equation}
where $\rho_i$ and $\theta_i$ are the density and phase of the $i$-th condensate. When $J_{ij} = G_{ij}$, since all the elements of the Google matrix are positive, the condensation threshold is realised at $\theta_i = \theta_j$ with the $\cos(\theta_j - \theta_i) = 1$ for all $i,j$. By denoting $x_i = \sqrt{\rho_i}$, we rewrite the real part of Eqs.~(\ref{eq:gd_cond_real_and_imag}) as:
\begin{equation}
  \frac{dx_i}{dt} = - x_i^3 + \gamma x_i + \sum_{j = 1}^N G_{ij} x_j.
  \label{eq:gd_cond_real}
\end{equation}
This time evolution of polariton and photon condensate amplitudes is similar to the networks of OPOs, described by Eq.~(\ref{eq:opo_networks}). As the effective pumping rate $\gamma$ increases from the negative values (linear dissipation dominates), the first nonzero stable state emerges at
\begin{equation}
  \gamma_{\rm crit} = - \lambda_{\rm max}^{({\bf G})} = -1,
\end{equation}
which corresponds to $\xi_{\rm crit} = \gamma_{\rm crit}$ in Eq.~(\ref{eq:general_network_linear}). Hence, the networks of gain-dissipative condensates emulate the PageRank algorithm in the regime of the lowest gain while the PageRank vector is represented by the absolute values of amplitudes $|\psi_i|$ at the steady state of equations:
\begin{equation}
  \frac{d \psi_i}{dt} = -i U |\psi_i|^2 \psi_i - (1 + |\psi_i|^2)\psi_i + \sum_{j = 1}^N G_{ij} \psi_j.
  \label{eq:gd_condensates_pagerank}
\end{equation}
In case of the Google matrix, the nonlinear interactions $U$ do not affect the dynamics of gain-dissipative condensates. Consequently, the dynamics of Eq.~(\ref{eq:gd_condensates_pagerank}) is equivalent to that of the networks of coupled lasers, which are considered next.

\textbf{Networks of lasers.}

The network of the degenerate lasers in a cavity represents a gain-dissipative unconventional computing hardware that was proposed for the minimisation of the XY Hamiltonian \cite{gershenzon2020exact}. In such networks, the interactions are engineered by mutual light injections from one laser to another, which introduce losses depending on the relative phases of lasers. The dynamics of coupled lasers is governed by the rate equations \cite{rogister2004power}:
\begin{equation}
  \begin{aligned}
       \frac{dE_i}{dt} &= (\widetilde{G}_i - \widetilde{\alpha}) E_i + \sum_{j=1}^N J_{ij} E_j, \\
       \frac{d \widetilde{G}_i}{dt} &= \frac{1}{\tau} [\widetilde{P} - \widetilde{G}_i (1 + |E_i|^2)],
       \label{eq:gd_lasers_system}
  \end{aligned}
\end{equation}
where $E_i$ is the electric field of the $i$-th laser, $\widetilde{G}_i$ is the active medium gain, $\tau$ is the gain medium fluorescence lifetime, $\widetilde{\alpha}$ is the linear loss coefficient, and $\widetilde{P}$ is the active medium pump rate, $J_{ij}$ are the coupling strengths between the $i$-th and $j$-th lasers. In the limit of the fast active medium gain relaxation and low amplitude electric fields, the equations simplify to:
\begin{equation}
  \frac{dE_i}{dt} = (\widetilde{P} - \widetilde{\alpha} - \widetilde{P} |E_i|^2) E_i + \sum_{j=1}^N J_{ij} E_j.
  \label{eq:gd_lasers}
\end{equation}
Consequently, the stability analysis of the dynamics of polariton and photon condensates, governed by Eq.~(\ref{eq:gd_condensates}), applies to the time evolution of coupled laser oscillators described by Eq.~(\ref{eq:gd_lasers}).
For the Google matrix, the first nonzero stable state occurs when the linear losses $\widetilde{\alpha}$ satisfy
\begin{equation}
  \widetilde{P} - \widetilde{\alpha}_{\rm crit} =  - \lambda_{\rm max}^{({\bf G})} = -1.
\end{equation}
The dynamics of linearised laser networks is equivalent to Eq.~(\ref{eq:general_network_linear}) at $\xi_{\rm crit} = \widetilde{P} - \widetilde{\alpha}_{\rm crit}$. Hence, the networks of coupled lasers emulate the PageRank algorithm in the lowest loss regime with the PageRank vector represented by the absolute values of electric fields $|E_i|$ in the steady state of equations:
\begin{equation}
  \frac{dE_i}{dt} = - (1 + \widetilde{P} |E_i|^2) E_i + \sum_{j=1}^N G_{ij} E_j.
  \label{eq:gd_lasers_pagerange}
\end{equation}
The dynamic of laser networks is reminiscent of gain-dissipative condensates described by Eq.~(\ref{eq:gd_condensates_pagerank}), where the presence of nonlinear term $U$ does not affect the system's ability to find the principal eigenvector. Consequently, the emulation of the PageRank algorithm with networks based on either polariton and photon condensates or lasers can be performed with Eq.~(\ref{eq:gd_condensates_pagerank}).

\textbf{Hopfield neural networks.}

The Hopfield networks \cite{hopfield1982neural} can be realised with unconventional hardware based on electronic \cite{cai2020power} and photonic systems \cite{tait2017neuromorphic} with quantum extensions available \cite{rebentrost2018quantum}. These networks are of great importance in many areas with early applications ranging from minimising discrete spin Hamiltonians and associative memory \cite{Hopfield1985} to more recent uses in web information retrieval \cite{chau2007incorporating}, pattern recognition \cite{krotov2016dense}, and natural language processing techniques \cite{ramsauer2020hopfield}. The evolution of individual neurons is governed by the equations:
\begin{equation}
  \frac{dx_i}{dt} = - \frac{x_i}{\tau} + \sum_{j=1}^N J_{ij} \tanh \big( \frac{x_j}{u_0} \big) + I_i^b,
  \label{eq:hopfield-tank}
\end{equation}
where $x_i$ describes the state of the $i$-th neuron, $\tau$ is the leakage parameter, $J_{ij}$ are the interaction coefficients between neurons,  $I_i^b$ is the external pumping imposed on the $i$-th neuron, and the activation function is assumed to be the hyperbolic tangent. In the case of the Google matrix, the first nonzero stable state emerges at
\begin{equation}
 \tau_{\rm crit} = \frac{u_0}{\lambda_{\rm max}^{({\bf G})}} = u_0.
\end{equation}
The dynamics of the linearised Hopfield networks is equivalent to such of the Eq.~(\ref{eq:general_network_linear}) at $\xi_{\rm crit} = - 1 / \tau_{\rm crit}$. Hence, the Hopfield networks emulate the PageRank algorithm in the lowest loss regime with the PageRank vector represented by the amplitudes $x_i$ in the steady-state of equations:
\begin{equation}
  \frac{dx_i}{dt} = - x_i + \sum_{j = 1}^N G_{ij} \tanh(x_j) + I_i^b,
  \label{eq:hopfield-tank_pagerank}
\end{equation}
where the time and amplitudes in Eq.~(\ref{eq:hopfield-tank}) are rescaled as $t \rightarrow t / u_0$ and $x_i \rightarrow x_i / u_0$.

\subsection*{4. Numerical parameters}

The numerical results presented in the main text of the article are achieved for the OPO-networks by simulating Eq.~(\ref{eq:opo_networks_pagerank}) with $\beta = 1$, polariton/photon/laser networks by simulating Eq.~(\ref{eq:gd_condensates_pagerank}) with $U = 1$, Hopfield networks by simulating Eq.~(\ref{eq:hopfield-tank_pagerank}) with $I_i^b = 0$ for all elements. The Euler iterative scheme is used for all networks with the time step $dt = 1$. We note that smaller time steps would work too, and the choice of such large $dt$ makes the time evolution of the considered networks similar to the power method, albeit in the presence of small nonlinearities, which could affect the components of the PageRank vectors for some datasets. In Fig.~\ref{Fig2} and Fig.~\ref{Fig3}, the L1-norm termination criterion is used for all algorithms with the maximum number of iterations corresponding to the $10^{-16}$ tolerance.

For the numerical calculations of the XYRank, the numerical parameters in Eqs.~(\ref{eq:gd_xy_model_psi}, \ref{eq:gd_xy_model_gamma}) are fixed to be $\rho_{\rm th} = 10$, $\epsilon = 60$, $dt = 0.005$, across datasets \textit{`harvard'}, \textit{`california'}, and \textit{`facebook'}, while for the \textit{`wiki-topcats'} dataset $\rho_{\rm th} = 1$, $\epsilon = 15$, $dt = 0.0005$. The presented rankings in Table~\ref{table:pr_top20} and Table~\ref{table:pr_top20_reverse} are consistent across different choices of parameters with gain-dissipative networks converging to the similar steady state under the fixed tolerance, although the required number of iterations for convergence greatly depends on a particular choice. The Euler method with L1 norm accuracy of $10^{-10}$ converges in about 5000, 4000, and 50000 iterations for datasets \textit{`harvard'}, \textit{`california'}, and \textit{`facebook'}, while the maximum limit of 1000000 iterations is reached for the \textit{`wiki-topcats'} dataset. The easiness of the Google matrices for the minimisation of the XY model is caused by the ferromagnetic sign of all couplings.

\subsection*{5. The highest XYRank positions}

The XYRank is an alternative ranking that is calculated through the minimisation of the XY Hamiltonian on the Google matrix. Similar to the highest PageRank positions in Table~\ref{table:pr_top20}, the highest XYRank positions are shown in Table~\ref{table:pr_top20_reverse}. The Kendall's tau between \mbox{PageRank} and XYRank across all positions is 0.74, 0.16, 0.74, 0.61, for the datasets \textit{`harvard'}, \textit{`california'}, \textit{`facebook'}, and \textit{`wiki-topcats'}, respectively.
\begin{table*}[t]
  \begin{ruledtabular}
    \caption{The highest 10 alternative ranking positions (XYRank) are shown for datasets \textit{`harvard'}, \textit{`california'}, \textit{`facebook'}, and \textit{`wiki-topcats'} of size $N = 500$, $N = 9664$, $N = 22470$, and $N = 1791489$. The XYRank is calculated by minimising the XY Hamiltonian for the Google matrices with Eqs.~(\ref{eq:gd_xy_model_psi}-\ref{eq:gd_xy_model_gamma}). The difference between the XYRank and PageRank distributions is indicated by green (red) arrows showing the shift in the PageRank towards a higher (lower) rating by a certain number of positions with respect to the XYRank.
     }
    \centering
    \label{table:pr_top20_reverse}
    \begin{tabular}{c l l l l}
      XYRank & Harvard & PageRank & California & PageRank \\
      \hline
      1 & www.harvard.edu & $-$ & www.linkexchange.com/ & $\color{red}\Downarrow5$ \\
			2 & search.harvard.edu:8765/custom/.. & $\color{red}\Downarrow1$ & home.netscape.com/comprod/mirror/.. & $\color{red}\Downarrow2$ \\
			3 & www.med.harvard.edu & $\color{red}\Downarrow1$ & www.yahoo.com/ & $\color{red}\Downarrow9$ \\
			4 & www.hsph.harvard.edu & $\color{red}\Downarrow4$ & www.berkeley.edu/ & $\color{red}\Downarrow3$ \\
			5 & www.gse.harvard.edu & $-$ & www.berkeley.edu & $-$ \\
			6 & www.ksg.harvard.edu & $\color{red}\Downarrow1$ & www.leginfo.ca.gov/calaw.html & $\color{red}\Downarrow20$ \\
			7 & search.harvard.edu:8765/query.html & $\color{red}\Downarrow4$ & www.creia.com/ & $\color{red}\Downarrow25$ \\
			8 & www.hbs.edu & $\color{green}\Uparrow6$ & www.ca.gov/ & $\color{red}\Downarrow11$ \\
			9 & www.hms.harvard.edu & $\color{green}\Uparrow3$ & www.adobe.com/prodindex/acrobat/.. & $\color{red}\Downarrow15$ \\
			10 & www.gse.harvard.edu/search.html & $\color{red}\Downarrow7$ & www.dot.ca.gov/hq/roadinfo/i80 & $\color{red}\Downarrow24$ \\
      \hline
      XYRank & Facebook & PageRank & Wikipedia (top categories) & PageRank \\
      \hline
			1 & Sir Peter Bottomley MP & $\color{red}\Downarrow1$ & United States & $-$ \\
			2 & Joachim Herrmann & $\color{red}\Downarrow5$ & France & $-$ \\
			3 & Facebook & $\color{green}\Uparrow2$ & Departments of France & $\color{red}\Downarrow12$ \\
			4 & Harish Rawat & $\color{red}\Downarrow7$ & Communes of France & $\color{red}\Downarrow14$ \\
			5 & U.S. Army Chaplain Corps & $\color{red}\Downarrow1$ & Canada & $\color{green}\Uparrow1$ \\
			6 & Manfred Weber & $\color{red}\Downarrow4$ & Village & $\color{red}\Downarrow18$ \\
			7 & Home $\&$ Family & $\color{red}\Downarrow9$ & Powiat & $\color{red}\Downarrow18$ \\
			8 & Cancillería Argentina & $\color{red}\Downarrow18$ & Gmina & $\color{red}\Downarrow14$ \\
			9 & Barack Obama & $\color{green}\Uparrow1$ & Voivodeships of Poland & $\color{red}\Downarrow14$ \\
			10 & Loïc Hervé & $\color{red}\Downarrow15$ & Germany & $\color{green}\Uparrow5$ \\
    \end{tabular}
  \end{ruledtabular}
\end{table*}

\subsection*{6. Web graph datasets}

The considered datasets are publicly available and include:
\begin{itemize}
  \item \textit{Harvard web graph (`harvard').} The Harvard database is an $N = 500$ directed graph containing web pages related to Harvard University as of 2002  \cite{moler2004numerical,harvard_database}.
  \item \textit{California web graph (`california').} The California database is an $N = 9, 664$ directed graph that contains pages matching the query ``California" collected in 2002 \cite{california_database}.
  \item \textit{Facebook web graph (`facebook').} The Facebook database is an $N = 22,470$ undirected graph with nodes representing official Facebook pages and edges corresponding to mutual likes between pages. Pages belong to 4 categories: politicians, governmental organisations, television shows, and companies. The graph is collected in 2017 and is available in Stanford large network dataset collection \cite{snapnets}.
  \item \textit{Cambridge University web graph (`cambridge').} The Cambridge University database is an $N = 212,710$ directed graph containing web pages related to the University of Cambridge as of 2006 \cite{cambridge_database}. The spectral properties of graphs of different UK universities are analysed in
  \cite{frahm2011universal}.
  \item \textit{Amazon web graph (`amazon').} The Amazon database is an $N = 400,727$ directed graph collected in 2003 \cite{snapnets} with nodes representing products and edges corresponding to frequently co-purchased products.
  \item \textit{Hollywood web graph (`hollywood').} The Hollywood database is an $N = 1,139,905$ undirected graph collected in 2009 \cite{rossi2015network} with nodes representing actors and edges corresponding to the appearance of actors in the same movies.
  \item \textit{Wikipedia top categories web graph (`wiki-topcats').} The Wikipedia top categories database is an ${N = 1,791,489}$ directed graph collected in 2011 \cite{snapnets} with nodes representing pages from the top Wikipedia categories (have at least 100 pages) with the largest strongly connected component and edges corresponding to hyperlinks.
  \item \textit{Wikipedia web graph (`wikipedia').} The Wikipedia database is an $N = 3,566,907$ directed graph collected in 2007 \cite{wikipedia_database} with nodes representing Wikipedia pages and edges corresponding to hyperlinks between them.
\end{itemize}

All datasets can also be found on the GitHub page together with the implementations of numerical methods for calculating the PageRank.

\subsection*{7. Data for computing power and energy efficiency of classical devices}

 The numbers in Fig.~\ref{Fig4} are representative of typical orders of computing power and energy efficiency of contemporary classical computing architectures, although they could be drastically different within one class of computing devices. In the descriptions below, we denote the processing power by $R$ that is measured in the number of floating-point operations per second (${\rm FLOP/s = FLOPS}$), the power consumption by $P$ that is measured in watts (${\rm W}$), and energy efficiency is calculated as the ratio $R/P$ and is measured in units ${\rm FLOPS/W} =  {\rm [FLOP/J]}$. The performance for devices is reported for either double-precision (${\rm FP64}$), single-precision (${\rm FP32}$), half single-precision (${\rm FP16}$), or integer precision (${\rm int6}$) calculations.

{\begin{itemize}
  \item[1.] \textit{Supercomputers.} In terms of energy efficiency, the top 10 list of supercomputers \cite{top500} starts with
  \begin{center}
    \textit{NVIDIA DGX SuperPOD} (FP64): \\
    $R = 2356 \ {\rm TFLOP/s}$, $P = 90 \ {\rm kW}$, \\ $R/P = 26.2 \ {\rm GFLOPS/W}$
  \end{center}
  and ends with
  \begin{center}
    \textit{Fujitsu's Supercomputer Fugaku} (FP64): \\
    $R = 442000 \ {\rm TFLOP/s}$, $P = 29899 \ {\rm kW}$, \\ $R/P = 14.8 \ {\rm GFLOP/J}$.
  \end{center}
  The latter is also the most powerful supercomputer in terms of processing power, that has overcome the previous best supercomputer
  \begin{center}
     \textit{Summit IBM} (FP64): \\
     $R = 148600 \ {\rm TFLOP/s}$, $P = 10096 \ {\rm kW}$,\\ $R/P = 14.72 \ {\rm GFLOPS/W}$.
  \end{center}
  \item[2.] \textit{GPUs.} For an estimate of GPU power efficiency we consider two state-of-the-art cards \cite{nvidia}:
  \begin{center}
     \textit{NVIDIA GeForce RTX 3090} (FP64): \\
     $R = 0.556 \ {\rm TFLOP/s}$, $P = 0.35 \ {\rm kW}$,\\ $R/P = 1.85 \ {\rm GFLOPS/W}$. \\
     \textit{NVIDIA GeForce RTX 3090} (FP32): \\
     $R = 36 \ {\rm TFLOP/s}$, $P = 0.35 \ {\rm kW}$,\\ $R/P = 103 \ {\rm GFLOPS/W}$.
  \end{center}
  and
  \begin{center}
     \textit{NVIDIA V100} (FP64): \\
     $R = 7.8 \ {\rm TFLOP/s}$, $P = 0.3 \ {\rm kW}$,\\ $R/P = 26 \ {\rm GFLOPS/W}$ \\
     \textit{NVIDIA V100} (FP32): \\
     $R = 15.7 \ {\rm TFLOP/s}$, $P = 0.3 \ {\rm kW}$,\\ $R/P = 52.3 \ {\rm GFLOPS/W}$ \\
     \textit{NVIDIA V100} (FP16): \\
     $R = 125 \ {\rm TFLOP/s}$, $P = 0.3 \ {\rm kW}$,\\ $R/P = 417 \ {\rm GFLOPS/W}$.
  \end{center}
  \item[3.] \textit{CPUs.} Most CPUs lie within processing power of $2 \ {\rm TFLOP/s}$ and power efficiency of about $10 \ {\rm GFLOPS/W}$.  As an estimate of top CPU power efficiency we use \cite{sun2019summarizing}
  \begin{center}
     \textit{Intel Xeon} (FP64): \\
     $R = 4.8 \ {\rm TFLOP/s}$, $P = 0.165 \ {\rm kW}$,\\ $R/P = 29 \ {\rm GFLOPS/W}$ \\
     \textit{Intel Xeon} (FP32): \\
     $R = 9. \ {\rm TFLOP/s}$, $P = 0.165 \ {\rm kW}$,\\ $R/P = 55 \ {\rm GFLOPS/W}$
  \end{center}
  \item[4.] \textit{Dedicated hardware.} FPGAs are reprogrammable hardware devices that provide energy efficient computing tailored specific tasks. One of the high-end FPGA boards \cite{parker2017understanding,nurvitadhi2017can} is
  \begin{center}
     \textit{Intel Stratix10} (FP32): \\
     $R = 10 \ {\rm TFLOP/s}$, $P = 0.18 \ {\rm kW}$,\\ $R/P = 56 \ {\rm GFLOPS/W}$ \\
     \textit{Intel Stratix10} (int6): \\
     $R = 70. \ {\rm TFLOP/s}$, $P = 0.18 \ {\rm kW}$,\\ $R/P = 389 \ {\rm GFLOPS/W}$
  \end{center}
  The Tensor Processing Unit (TPU) is a custom application-specific integrated circuit (ASIC) designed by Google \cite{google_tpu} and used for accelerating machine learning tasks:
  \begin{center}
     \textit{TPU v3} (FP16): \\
     $R = 90. \ {\rm TFLOP/s}$, $P = 0.225 \ {\rm kW}$,\\ $R/P = 400 \ {\rm GFLOPS/W}$
  \end{center}
\end{itemize}}

Note that with distributed computing, when the processing power of personal computers is linked together over the Internet, the total computing power over 2.3 exaFLOP could be achieved as of 2020 \cite{beberg2009folding}.

As an estimate of the energy consumption of computing the PageRank vector of 10 billion size matrix in the main text, we have assumed that one needs to run 1000 iterations of the power method, and there are around 100 connections per each element in the matrix. The PageRank vector's single computation would then take around $1 \  {\rm PFLOP}$. To update the ranking on a minute scale, the PageRank vector would need to be recomputed about half a million times over a year. The average annual electricity consumption for computing the PageRank would be around $3.5 \cdot 10^5 \ {\rm kWh}$ on dedicated hardware, such as the tensor processing units (TPUs) with the energy efficiency of $400 {\rm GFLOPS/W}$. This amount of energy is equivalent to charging up 290 electric cars for one year under the assumption of an average size battery of $30 \ {\rm kWh}$ and 40 charges per year.

\subsection*{8. Agreement between PageRank algorithm and Unconventional networks}

\begin{figure}[t!]
	\centering
	\includegraphics[width=8.6cm]{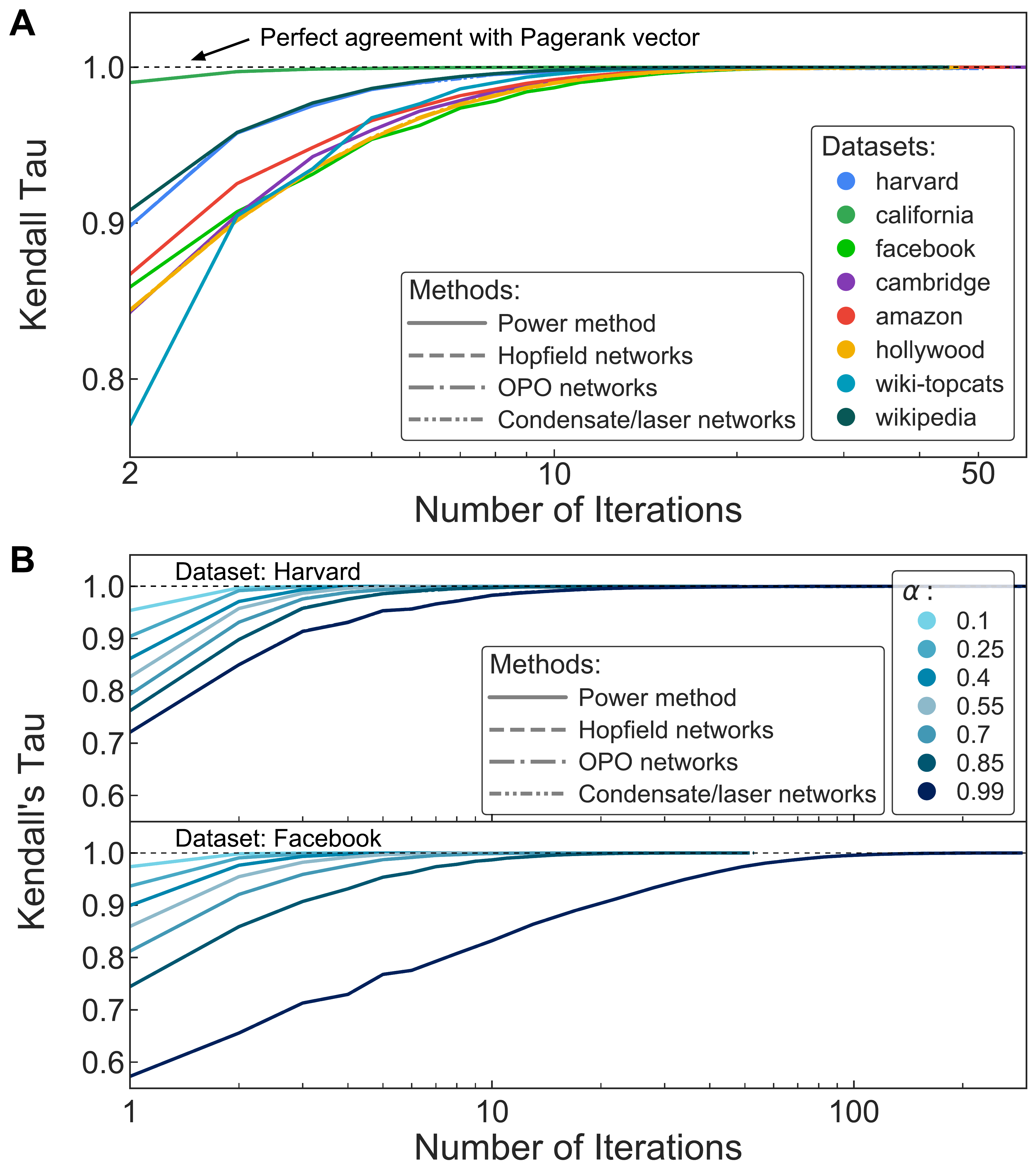}
	\caption{(A) Kendall's tau, as a measure of the agreement between the PageRank vector computed with the power method and the amplitude distributions obtained using unconventional networks, is shown as a function of the number of iterations for a variety of web graphs. The unconventional networks are represented by Hopfield networks and networks of optical parametric oscillators (OPO), condensates (polariton and photon), and lasers. The L1-norm termination criterion is used for all algorithms with the maximum number of iterations corresponding to the $10^{-6}$ tolerance. The teleportation parameter is fixed across all datasets to $\alpha = 0.85$. (B) Kendall's tau distribution is shown as a function of the number of iterations for a variety of teleportation parameter values for two datasets: Harvard network (top) and Facebook network (bottom). For all plots in (A) and (B) the dashed lines fully coincide with solid lines reflecting strong dynamics equivalence between unconventional networks and the power method.
  }
   \label{Fig_supplementary}
\end{figure}

The Kendall's tau between the PageRank vector, calculated with the power method, and the amplitude distributions of steady states, obtained with unconventional networks, demonstrates nearly perfect agreement across all datasets in Fig.~\ref{Fig_supplementary}A and identical convergence for all teleportation parameter values in Fig.~\ref{Fig_supplementary}B.

\section{Additional Information}
The authors declare that they have no competing interests.

\section{Keywords}
PageRank, Information Search and Retrieval, Unconventional Hardware, Gain-Dissipative systems, XY Hamiltonian, Optical Computing

\bibliographystyle{apsrev4-1}
\bibliography{Refs-pr2}

\end{document}